\documentclass{birkjour}
\newtheorem{thm}{Theorem}[section]
\newtheorem{cor}[thm]{Corollary}

\theoremstyle{defn}
\newtheorem{defn}[thm]{Definition}
\theoremstyle{rem}

\numberwithin{equation}{section}

\begin{document}

%
%
%
%
%
%
%
%
%

\title{A sharp Clifford-wavelet Heisenberg-type uncertainty principle}

\author[Banouh]{Hicham Banouh}
\address{Laboratoire de Syst\`emes Dynamiques et G\'eom\'etrie, Facult\'e de Math\'ematiques, Universit\'e de Sciences et Technologie Houari Boumedienne, Bab Zouar, Alger, Algeria.}
\email{hbanouh@usthb.dz}
\author[Ben Mabrouk]{Anouar Ben Mabrouk}
\address{Higher institut of Applied Mathematics and Computer Sciences, University of Kairaouan, Street of Assad Ibn Alfourat, Kairaouan 3100, Tunisia.\\
and\\
Laboratory of Algebra, Number Theory and Nonlinear Analysis LR18ES15, Department of Mathematics, Faculty of Sciences, Monastir, Tunisia.\\
and\\
Department of Mathematics, Faculty of Sciences, University of Tabuk, King Faisal Rd, Tabuk, Saudi Arabia.}
\email{anouar.benmabrouk@fsm.rnu.tn}

\begin{abstract} 
In the present work we are concerned with the development of a new uncertainty principle based on wavelet transform in the Clifford analysis/algebras framework. We precisely derive a sharp Heisenberg-type uncertainty principle for the continuous Clifford wavelet transform. 
\end{abstract}
\maketitle
\section{Introduction}
In the last decade(s) a coming back to Clifford algebras and analysis has taken place in many fields in mathematics, physiscs as well as computer science. This may be due to the fact that Clifford algebras incorporate inside one single structure the geometrical and algebraic properties of Euclidean space. For backgrounds on Clifford algebras/analysis the readers may refer to the original works  \cite{Clifford1882,Dirac1928,Grossmann1986,Grossmann1984,Hamilton1844,Hamilton1866,Weyl1950} which are already free.

Insite of a Clifford algebra geometric entities are treated ccording to their dimension such as scalars, vectors, bivectors and volume elements. This has allowed researchers especially in mathematical physics to develop harmonic analysis on Clifford algebras as extension of complex one.  Famous operators and related inequalities such as Cauchy-Riemann operator and uncertainty principle have been extended to the Clifford framework. 

Recently and especially with the research group of Ghent university (Sommen, Brackx, Delange, De Schepper, Hitzer, Mawardi, and their collaborators) have extended wavelet theory to the Clifford framework. Next, a series of works have been developed in the same topic. See \cite{Arfaoui-Rezgui-BenMabrouk,Arfaoui-BenMabrouk1,Arfaoui-BenMabrouk2,Arfaoui-BenMabrouk-Cattani1,Arfaoui-BenMabrouk-Cattani2,BrackxChisholmSoucek,Brackx-Schepper-Sommen0,Brackxetal2006,Brackx-Schepper-Sommen2,Brackxetal2006a,Brackxetal2013,Brackx2001a,Bujacketal,Bujacketal1,Bujacketal2,Delanghe,DeBie1,DeBie2,DeSchepper2006,Hitzer,Hitzer1,Hitzer2,Hitzer-Mawardi-1,Hitzer-Tachibana,Mawardi,Mawardi-Hitzer-1,Mawardi-Ashino-Vaillancourt,Sommen2015,Wietzke,Zou}. 

In the original theory of wavelets on $\mathbb{R}$ as well as its extension to other more general contexts, wavelet analysis of functions starts by checking that the mother wavelet is admissible or not to guarantee the most useful rules by the next such as the reconstruction, Fourier-Plancherel and Parseval ones. For real wavelet analysis, this is explained as follows. Given an analyzing function $\psi\in L^2(\mathbb{R})$ which will be called by the next the mother wavelet, such a function $\psi$ has to satisfy the admissibility condition  
\[
\mathcal{A}_{\psi}=\int_{\mathbb{R}}\frac{\left|\widehat{\psi}(\xi)\right|^{2}}{\left|\xi\right|}d\xi<+\infty,
\]
where $\widehat{\psi}$ is the classical Fourier transform of $\psi$.

The present work is concerned with the development of a sharp uncertainty principle in the Clifford framework based on Clifford wavelets and thus extend the one developed by Banouh et al in \cite{Banouh}. 

As we have mentioned above, the connection between wavelet theory and Clifford algebra/analysis is now growing up and applications are widespread such as in signal/image processing. See \cite{Brackx2001a,DeSchepper2006,Sommen2015,Rizo}. Applications also in physics, nuclear magnetics, electric engineering, color images may be found in \cite{Dian,Hitzer,Rizo,Sauetal,Soulard,Wietzke}.

The aim of the present paper is to establish a new Heisenberg uncertainty principle
applied to continuous Clifford wavelet transform in the settings of the non commutative Clifford algebras. In section 2 some preliminaries on Clifford toolkits and wavelets are presented. Section 3 is concerned to a review on the uncertainty principle in both its classical forms and Clifford one. Section 4 is devoted to the development of our main results. 
\section{Clifford wavelets toolkit}
In this section, we review briefly the basic concepts of Clifford analysis and the continuous wavelet transform on the real Clifford algebra. 
\subsection{Clifford analysis/algebra toolkit}
The Clifford algebra $\mathbb{R}_{n}$ associated to $\mathbb{R}^{n}$ is the $2^n$-dimensional non commutative associative algebra generated by the canonical basis $\left\{e_{1},e_{2},\dots,e_{n}\right\} $ relatively to the non commutative product
\[
e_{i}e_{j}+e_{j}e_{i}=-2\delta_{ij},
\]
where $\delta$ is the Kronecker symbol. Any element $a\in\mathbb{R}_{n}$ has a unique representation 
\[
a=\sum_{A}a_{A}e_{A},\;a_{A}\in\mathbb{R}=\sum_{k=0}^{n}\sum_{\left|A\right|=k}a_{A}e_{A},
\]
where $|A|$ is the length of the multi-index $A$.\\
On the algebra $\mathbb{R}_{n}$ there are some types of involutive operators defined on the basis elements. A main-involution satisfying $\widetilde{e_{A}}=(-1)^{\left|A\right|}e_{A}$. A second type called reversion stating that $e_{A}^{\ast}=(-1)^{\frac{\left|A\right|(\left|A\right|-1)}{2}}e_{A}$ and finally a conjugation operator defined by $\overline{e_{A}}=(-1)^{\frac{|A|(|A|+1)}{2}}e_{A}$. In the case of the complex extension $\mathbb{C}_{n}=\mathbb{R}_{n}+i\mathbb{R}_{n}$ the last operator has to be extended to the so-called Hermitian conjugation defined by $\lambda^{\dagger}=\overline{a}-i\overline{b}$ for $\lambda=a+ib\in\mathbb{C}_{n}$ with $a,b\in\mathbb{R}_{n}$. 

As the vextor space $\mathbb{R}^{n}$ is now seen as a subspace of $\mathbb{R}_{n}$ any element $x=(x_{1},x_{2},\dots,x_{n})\in\mathbb{R}^{n}$
is also seen as an element in $\mathbb{R}_{n}$. We precisely identify it to $\underline{x}={\displaystyle \sum_{j=1}^{n}x_{j}e_{j}}$. This allows to introduce the Clifford product of vectors by
$$
\underline{x}\underline{y}=\underline{x}\cdot\underline{y}+\underline{x}\wedge\underline{y},
$$
where 
$$
\underline{x}\cdot\underline{y}=-<\underline{x},\underline{y}>=-{\displaystyle \sum_{j=1}^{n}x_{j}y_{j}},
$$
and 
$$
\underline{x}\wedge\underline{y}={\displaystyle\sum_{j<k}e_{j}e_{k}(x_{j}y_{k}-x_{k}y_{j})}.
$$
This yields particularly that 
$$
\underline{x}^{2}=-\left|\underline{x}\right|^{2}=-\sum_{j=1}^{n}|x_{j}|^{2}.
$$
Finally we recall that for two functions $f,g:\mathbb{R}^n\longrightarrow\mathbb{C}_n$ the inner product is defined by
\begin{equation}\label{eq:inner product}
<f,g>_{L^{2}(\mathbb{R}_{n},dV(\underline{x}))} =\int_{\mathbb{R}^{n}}\left[f(\underline{x})\right]^{\dagger}g(\underline{x})dV(\underline{x})
\end{equation}
and the associated norm by
$$
\|f\|_{L^{2}(\mathbb{R}_{n},dV(\underline{x}))} =<f,f>_{L^{2}(\mathbb{R}_{n},dV(\underline{x}))}^{\frac{1}{2}}.
$$
We denote also
\[
\left|f\right|_{L^{1}(\mathbb{R}_{n},dV(\underline{x}))}=\int_{\mathbb{R}^{n}}|f(\underline{x})|dV(\underline{x})
\]
where $dV(\underline{x})$ stands for the Lebesgue measure. As in the classical cases, it is easy to show the Cauchy-Schwartz inequality
\begin{equation}
\left|<f,g>_{L^{2}(\mathbb{R}_{n},dV(\underline{x}))}\right|\leq\left|f\right|_{L^{2}(\mathbb{R}_{n},dV(\underline{x}))}\left|g\right|_{L^{2}(\mathbb{R}_{n},dV(\underline{x}))}.\label{cauchy schwartz}
\end{equation}
\subsection{The continuous Clifford-wavelet transform}
As for multidimensional Euclidean spaces the Fourier and wavelet transforms have been extended to the case of Clifford framework. The Clifford-Fourier transform of a Clifford-valued function $f\in L^{1}\cap L^{2}(\mathbb{R}_{n},dV(\underline{x}))$
is defined by
\[
\mathcal{F}\left[f\right](\underline{\xi})=\widehat{f}(\underline{\xi})=\frac{1}{(2\pi)^{\frac{n}{2}}}\int_{\mathbb{R}^{n}}e^{-i<\underline{x},\underline{\xi}>}f(\underline{x})dV(\underline{x}).
\]
Here also we may speak about its inverse defined by 
\[
\mathcal{F}^{-1}\left[f\right](\underline{x})=\frac{1}{(2\pi)^{\frac{n}{2}}}\int_{\mathbb{R}^{n}}e^{i<\underline{x},\underline{\xi}>}\widehat{f}(\underline{\xi})dV(\underline{\xi}).
\]
The concept of Clifford-wavelet transform has been now widely studied. In the present paper we will recall a special case based on the spin theory which will be used in our work and which has been already applied in \cite{Banouh}. Let $\psi\in L^{1}\cap L^{2}(\mathbb{R}^{n},dV(\underline{x}))$ be such that 
\begin{itemize}
\item ${\psi}$ is a Clifford-algebra-valued function.
\item $\widehat{\psi}(\underline{\xi})\left[\widehat{\psi}(\underline{\xi})\right]^{\dagger}$ is scalar.
\item $\mathcal{A}_{\psi}=\displaystyle(2\pi)^{n}\int_{\mathbb{R}^{n}}\frac{\widehat{\psi}(\underline{\xi})\left[\widehat{\psi}(\underline{\xi})\right]^{\dagger}}{|\underline{\xi}|^{n}}dV(\underline{\xi})<\infty.$
\end{itemize}
$\psi$ is called by the next a Clifford mother wavelet. We also recall in the context of Clifford framework, the spin group of order $n$ which is defined by
\[
Spin(n)=\left\{ s\in\mathbb{R}_{n};\;s={\displaystyle \prod\limits _{j=1}^{2l}}\underline{\omega}_{j},\;\underline{\omega}_{j}^{2}=-1,1\leq j\leq2l\right\}.
\]
For $(a,\underline{b},s)\in\mathbb{R}_{+}\times\mathbb{R}^{n}\times Spin(n)$,
we denote
\[
\psi^{a,\underline{b},s}(\underline{x})=\frac{1}{a^{\frac{n}{2}}}s\psi(\frac{\overline{s}(\underline{x}-\underline{b})s}{a})\overline{s}.
\]
In \cite{Banouh} the quthors showed that the copies ${\psi^{a,\underline{b},s}}$ and that they constitute a dense set in $L^{2}(\mathbb{R}^{n},dV(\underline{x}))$.
\begin{defn} The Clifford-wavelet transform of an analyzed function $f\in L^2(\mathbb{R}^n,dV(\underline{x}))$ is defined by
	\[
	T_{\psi}\left[f\right](a,\underline{b},s)=\int_{\mathbb{R}^{n}}\left[\psi^{a,\underline{b},s}(\underline{x})\right]^{\dagger}f(\underline{x})dV(\underline{x}).
	\]
Let $\mathcal{H}_{\psi}=T_{\psi}\left(L^{2}(\mathbb{R}^{n},dV(\underline{x}))\right)$. We define an inner product on $\mathcal{H}_{\psi}$ by
\[
\left[T_{\psi}\left[f\right],T_{\psi}\left[g\right]\right]=\frac{1}{\mathcal{A}_{\psi}}\int\limits_{Spin(n)}\int\limits_{\mathbb{R}^{n}}\int\limits _{\mathbb{R}_{+}}(T_{\psi}\left[f\right](a,\underline{b},s))^{\dagger}T_{\psi}\left[g\right](a,\underline{b},s)d\mu(a,\underline{b},s),
\]
where $d\mu(a,\underline{b},s)=\frac{da}{a^{n+1}}dV(\underline{b})ds$ and where $ds$ is the Haar measure on $Spin(n)$.
\end{defn}
It is straightforwrd that the operator $T_{\psi}$ is an isometry from  $L^{2}(\mathbb{R}^{n},dV(\underline{x}))$ to $\mathcal{H}_{\psi}$. See \cite{Banouh}. As a result any analyzed function $f\in L^{2}(\mathbb{R}^{n},dV(\underline{x}))$ may be reconstructed in the $L^2$-sense as 
\[
f(\underline{x})=\frac{1}{\mathcal{A}_{\psi}}\int\limits _{Spin(n)}\int\limits _{\mathbb{R}^{n}}\int\limits _{\mathbb{R}_{+}}\psi^{a,\underline{b},s}(\underline{x})T_{\psi}\left[f\right](a,\underline{b},s)\frac{da}{a^{n+1}}dV(\underline{b})ds.
\]
This constitutes the well-known Plancheral-Parsevall formula extended to the Clifford wavelet framework. 
\subsection{The uncertainty principle revisited}
The uncertainty principle is originally due to Heisenberg (See \cite{Heisenberg1927}, \cite{Heisenberg1985}) and constitutes since its discovery an interesting concepts in quantum mechanics. Mathematically, it is resumed by the fact that a non-zero function and its Fourier transform cannot both be sharply localized. 

In the last decades especially of the last century many studies have been developed tackling the uncertainty principle. Many variants have been thus proposed using pure Fourier transform, pure wavelets, Clifford Fourier transform and recently Clifford wavelet transform. In \cite{Amri-Rachdi-1} for example a Stein–Weiss type uncertainty inequality has been proved. In \cite{Amri-Rachdi-2} Fourier transform combined with Riemann–Liouville operator has been applied for a Hausdorff–Young inequality to yield an entropy based uncertainty principle and a Heisenberg–Pauli–Weyl inequality. In \cite{Rachdi-Meherzi} continuous wavelet transform has been applied to prove an analogue of Heisenberg's inequality (See also \cite{Rachdi-Amri-Hammami}, \cite{Rachdi-Herch}).

In \cite{Feichtinger-Grochenig} using Heisenberg group techniques, stable iterative algorithms for signal analysis and synthesis the authors established a variant of the Heisenberg uncertainty inequality. In \cite{Dahkleteal} an uncertainty principle has been shown using shearlet transform. El-Haoui et al in \cite{ElHaouietal} applied a quaternionic Fourier transform to derive an uncertainty principles including Heisenberg-Weyls, Hardys, Beurlings and logarithmic ones. See also \cite{ElHaoui-Fahlaoui}.

Next, in the Clifford framework Hitzer and collaborators were the most active group in developing uncertainty principle variants such as \cite{Hitzer1,Hitzer2} where some generalized Clifford wavelet uncertainty principles have been proposed. See also \cite{Hitzer-Tachibana}. Hitzer and Mawardi in \cite{Hitzer-Mawardi-1} extended the Fourier transform to some general Clifford geometric algebras leading to a a new uncertainty principle for multivector functions in the new Clifford geometric algebras. See also \cite{Mawardi-Ryuichi}, \cite{Mawardi-Ryuichi-1}, \cite{Mawardi-Ryuichi-2}, \cite{Mawardi-Hitzer-1}, \cite{Mawardi-Hitzer-2}, \cite{Mawardi-Hitzer-3}, \cite{Mawardi-Hitzer-4}, \cite{Mawardi-Hitzer-Hayashi-Ashino}, \cite{Mawardi-Ashino-Vaillancourt}.

In \cite{Ma-Zhao} the authors used quaternion ridgelet transform and curvelet transform associated to the quaternion Fourier transform to establish an uncertainty principle. Kou et al in \cite{Kouetal} applied the linear quaternion canonical transform has been applied to establish an uncertainty principle. Yang et al in \cite{Yangetal1} have investigated a stronger directional uncertainty principles based on Fourier transform.

For about the uncertainty principle and different variants based on Fourier and wavelet transforms may be found in \cite{Jorgensen}, \cite{Nagata}, \cite{Sen}, \cite{Stabnikov}, \cite{Yang-Kou}.

We propose by the next to recall some mathematical formulations of the uncertainty principle from both classical and recent forms. 
\begin{thm}\label{uncertainty-cas-general} (\cite{Banouh,Weyl1950}) Let $A$ and $B$ be two self-adjoint operators on a Hilbert space $X$ with domains $\mathcal{D}(A)$ and $\mathcal{D}(B)$ respectively and consider their commutator $\left[A,B\right]=AB-BA$. Then
\begin{equation}
\left|Af\right|_{2}\left|Bf\right|_{2}\geq\displaystyle\frac{1}{2}\left|<\left[A,B\right]f,f>\right|,\forall f\in\mathcal{D}(\left[A,B\right]).
\end{equation}
\end{thm}
Consider next the special case 
$$
A_{k}f(\underline{x})=x_{k}f(\underline{x})\;\;\hbox{and}\;\;
B_{k}f(\underline{x})=\partial_{x_{k}}f(\underline{x}),\;k=1,2,\cdots,n.
$$
By applying Theorem \ref{uncertainty-cas-general} above, we get the following result.
\begin{cor}\label{cor1}(\cite{Banouh}) 
\[
\left|A_{k}f\right|_{L^{2}(\mathbb{R}_{n},dV(\underline{x}))}\left|B_{k}f\right|_{L^{2}(\mathbb{R}_{n},dV(\underline{x}))}\geq\frac{1}{2}\left|<\left[A_{k},B_{k}\right]f,f>\right\vert.
\]
Furthermore,
\begin{equation}\label{uncer clifford-fourier}
\|x_{k}f\|_{L^{2}(\mathbb{R}_{n},dV(\underline{x}))}\|\xi_{k}\widehat{f}\|_{L^{2}(\mathbb{R}_{n},dV(\underline{x}))}\geq\frac{1}{2}\|f\|_{L^{2}(\mathbb{R}_{n},dV(\underline{x}))}^{2}.
\end{equation}
\end{cor}
Moreover, by considering the oerator $T_{\psi}$ the authors is \cite{Banouh} established the follozing result.
\begin{thm}\label{main-result-Banouh}(\cite{Banouh}) Let be $\psi\in L^{2}(\mathbb{R}^{n},\mathbb{R}_{n},dV(\underline{x}))$ be an admissible Clifford mother wavelet. Then for $f\in L^{2}(\mathbb{R}^{n},\mathbb{R}_{n},dV(\underline{x}))$. It holds for $1\leq k\leq n$ that 
\begin{equation}
{\displaystyle \left({\displaystyle \int_{Spin(n)}{\displaystyle \int_{\mathbb{R}^{+}}\left|b_{k}T\left[f\right](a,\cdot,s)\right|_{2}^{2}\frac{da}{a^{n+1}}ds}}\right)^{\frac{1}{2}}\left|\xi_{k}\widehat{f}\right|_{2}\geq\frac{(2\pi)^{\frac{n}{2}}}{2}\sqrt{A_{\psi}}\left|f\right|_{2}^{2}.}
\end{equation}
\end{thm} 
\section{A sharper Clifford wavelet uncertainty principle}
In the present section we state and prove our main result in the present work. It consists a sharper formulation of the Clifford-wavelet uncertainty principle. The main result is subject of the following theorem.
\begin{thm}\label{new-main-result}
Let $\psi\in L^{2}(\mathbb{R}^{n},\mathbb{R}_{n},dV(\underline{x}))$ be an admissible Clifford mother wavelet. Then for $f\in L^{2}(\mathbb{R}^{n},\mathbb{R}_{n},dV(\underline{x}))$, the following uncertainty principle holds,
\[
\begin{array}{ccc}
\left(\displaystyle\int_{Spin(n)}\int_{\mathbb{R}^{+}}\left\|b_{k}T_{\psi}\left[f\right](a,\cdot,s)\right\|^{2}\frac{da}{a^{n+1}}ds\right)^{\frac{1}{2}}\left\|\xi_{k}\widehat{f}\right\Vert\\
\\
\qquad\qquad\qquad\qquad\qquad\qquad\qquad \geq\sqrt{2^{n+1}\pi{}^{n}A_{\psi}}\left\{\left\|f\right\|^{2}+2\left|\left\langle f_1,f_2\right\rangle \right|\right\}
\end{array}
\]
where
$$
f_1(\underline{x})=\frac{1}{A_{\psi}}\int_{Spin(n)}\int_{\mathbb{R}^{n}}\int_{\mathbb{R}^{+}}\psi^{a,\underline{b},s}(\underline{x})\partial_{b_{k}}T_{\psi}\left[f\right](a,\underline{b},s)\frac{da}{a^{n+1}}dV(\underline{b})ds
$$
and
$$
f_2(\underline{x})=\frac{1}{A_{\psi}}\int_{Spin(n)}\int_{\mathbb{R}^{n}}\int_{\mathbb{R}^{+}}\psi^{a,\underline{b},s}(\underline{x})b_{k}T_{\psi}\left[f\right](a,\underline{b},s)\frac{da}{a^{n+1}}dV(\underline{b})ds.
$$
\end{thm}
\noindent\textbf{Proof.} Observe firstly that 
\[
\left\|x_{k}f\right\|\left\|\xi_{k}\widehat{f}\right\|\geq\sqrt{2}(\left\|f\right\|^{2}+\left|2\left\langle x_{k}\partial_{x_{k}}f,f\right\rangle \right|).
\]
We substitute $f(\bullet)$ by $T_{\psi}\left[f\right](a,\bullet,s)$. We obtain
$$
\begin{array}{lll}
\left\|b_{k}T_{\psi}\left[f\right](a,\bullet,s)\right\|\left\|\xi_{k}\widehat{T_{\psi}\left[f\right](a,\bullet,s)}\right\|\\ \qquad\qquad\qquad\qquad\qquad\geq\sqrt{2}(\left\|T_{\psi}\left[f\right](a,\bullet,s)\right\|^{2}+\mathcal{S}_{\psi}\left[f\right](a,\bullet,s))
\end{array}
$$
where 
$$
\mathcal{S}_{\psi}\left[f\right](a,\bullet,s)=\left|2\left\langle b_{k}\partial_{b_{k}}T_{\psi}\left[f\right](a,\bullet,s),T_{\psi}\left[f\right](a,\bullet,s)\right\rangle \right|.
$$
As a result,
$$
\begin{array}{lll}
&\displaystyle\int_{Spin(n)}\int_{\mathbb{R}^{+}}\left\|b_{k}T_{\psi}\left[f\right](a,\bullet,s)\right\|\left\|\xi_{k}\widehat{T_{\psi}\left[f\right](a,\bullet,s)}\right\|\frac{da}{a^{n+1}}ds\\
\\
&\quad\geq\sqrt{2}\displaystyle\int_{Spin(n)}\int_{\mathbb{R}^{+}}(\left\|T_{\psi}\left[f\right](a,\bullet,s)\right\|^{2}+\mathcal{S}_{\psi}\left[f\right](a,\bullet,s))\dfrac{da}{a^{n+1}}ds.
\end{array}
$$
By Cauchy-Schwartz we get
$$
\begin{array}{lll}
&\displaystyle\int_{Spin(n)}\int_{\mathbb{R}^{+}}\left\|b_{k}T_{\psi}\left[f\right](a,\bullet,s)\right\|\left\Vert \xi_{k}\widehat{T_{\psi}\left[f\right](a,\bullet,s)}\right\Vert \frac{da}{a^{n+1}}ds\\
\\
&\qquad\qquad\displaystyle\leq(\int_{Spin(n)}\int_{\mathbb{R}^{+}}\left\Vert b_{k}T_{\psi}\left[f\right](a,\bullet,s)\right\Vert ^{2}\dfrac{da}{a^{n+1}}ds)^{\frac{1}{2}}\\
\\
&\qquad\qquad\qquad\displaystyle\times(\int_{Spin(n)}\int_{\mathbb{R}^{+}}\left\Vert \xi_{k}\widehat{T_{\psi}\left[f\right](a,\bullet,s)}\right\Vert ^{2}\dfrac{da}{a^{n+1}}ds)^{\frac{1}{2}}
\end{array}
$$
Denote next 
$$
BT_{\psi}\left[f\right]=(\int_{Spin(n)}\int_{\mathbb{R}^{+}}\left\Vert b_{k}T_{\psi}\left[f\right](a,\bullet,s)\right\Vert ^{2}\frac{da}{a^{n+1}}ds)^{\frac{1}{2}}
$$
and
$$
\xi T_{\psi}\left[f\right]=(\int_{Spin(n)}\int_{\mathbb{R}^{+}}\left\Vert \xi_{k}\widehat{T_{\psi}\left[f\right](a,\bullet,s)}\right\Vert ^{2}\frac{da}{a^{n+1}}ds)^{\frac{1}{2}}.
$$
Then, we get 
\begin{align*}
BT_{\psi}\left[f\right]\times\xi T_{\psi}\left[f\right]
&\geq\sqrt{2}\int_{Spin(n)}\int_{\mathbb{R}^{+}}(\left\Vert T_{\psi}\left[f\right](a,\bullet,s)\right\Vert ^{2}\\
&\quad\quad+\left|2\left\langle b_{k}\partial_{b_{k}}T_{\psi}\left[f\right](a,\bullet,s),T_{\psi}\left[f\right](a,\bullet,s)\right\rangle \right|)\frac{da}{a^{n+1}}ds\\
&=\sqrt{2}\int_{Spin(n)}\int_{\mathbb{R}^{+}}(\left\Vert T_{\psi}\left[f\right](a,\bullet,s)\right\Vert ^{2}\frac{da}{a^{n+1}}ds\\
&\quad\quad+\sqrt{2}\int_{Spin(n)}\int_{\mathbb{R}^{+}}\mathcal{S}_{\psi}\left[f\right](a,\bullet,s)\frac{da}{a^{n+1}}ds.
\end{align*}
Observe next that 
$$
\int_{Spin(n)}\int_{\mathbb{R}^{+}}\left\Vert \xi_{k}\widehat{T_{\psi}\left[f\right]}(a,\bullet,s)\right\Vert ^{2}\frac{da}{a^{n+1}}ds=\frac{A_{\psi}}{(2\pi)^{n}}\left\Vert \xi_{k}\widehat{f}\right\Vert ^{2}
$$
and that 
$$
\int_{Spin(n)}\int_{\mathbb{R}^{+}}\left\Vert T_{\psi}\left[f\right](a,\bullet,s)\right\Vert ^{2}\frac{da}{a^{n+1}}ds =A_{\psi}\left\Vert f\right\Vert ^{2}
$$
and denote 
$$
\widetilde{BT_{\psi}}\left[f\right]=
\int_{Spin(n)}\int_{\mathbb{R}^{+}}\left|\left\langle b_{k}\partial_{b_{k}}T_{\psi}\left[f\right](a,\bullet,s),T_{\psi}\left[f\right](a,\bullet,s)\right\rangle \right|)\frac{da}{a^{n+1}}ds.
$$
We deduce that  
\begin{align*}
BT_{\psi}\left[f\right]\times(\frac{A_{\psi}}{(2\pi)^{n}}\left\Vert \xi_{k}\widehat{f}\right\Vert ^{2})^{\frac{1}{2}}\geq\sqrt{2}A_{\psi}\left\Vert f\right\Vert ^{2}+2\sqrt{2}\widetilde{BT_{\psi}}\left[f\right].
\end{align*}
Consequently,
$$
\begin{array}{lll}
&\displaystyle(\int_{Spin(n)}\int_{\mathbb{R}^{+}}\left\Vert b_{k}T_{\psi}\left[f\right](a,\bullet,s)\right\Vert ^{2}\frac{da}{a^{n+1}}ds)^{\frac{1}{2}}\left\Vert \xi_{k}\widehat{f}\right\Vert\\
\\
&\displaystyle\geq\sqrt{2^{n+1}\pi{}^{n}A_{\psi}}\left\Vert f\right\Vert^{2}
+\displaystyle\sqrt{\frac{2^{n+1}\pi{}^{n}}{A_{\psi}}}\int_{Spin(n)}\int_{\mathbb{R}^{+}}\mathcal{S}_{\psi}\left[f\right](a,\bullet,s)\dfrac{da}{a^{n+1}}ds.
\end{array}
$$
Now, observe that
$$
\begin{array}{lll}
\displaystyle\int_{Spin(n)}\int_{\mathbb{R}^{+}}\left|\left\langle b_{k}\partial_{b_{k}}T_{\psi}\left[f\right](a,\bullet,s),T_{\psi}\left[f\right](a,\bullet,s)\right\rangle \right|)\frac{da}{a^{n+1}}ds\\
\\
=\displaystyle\int_{Spin(n)}\int_{\mathbb{R}^{+}}\left|\int_{\mathbb{R}^{n}}\left[\partial_{b_{k}}T_{\psi}\left[f\right](a,\underline{b},s)\right]^{\dagger}b_{k}T_{\psi}\left[f\right](a,\underline{b},s)dV(\underline{b})\right|\frac{da}{a^{n+1}}ds\\
\\
\geq\displaystyle\left|\int_{Spin(n)}\int_{\mathbb{R}^{+}}\int_{\mathbb{R}^{n}}\left[\partial_{b_{k}}T_{\psi}\left[f\right](a,\underline{b},s)\right]^{\dagger}b_{k}T_{\psi}\left[f\right](a,\underline{b},s)dV(\underline{b})\frac{da}{a^{n+1}}ds\right|\\
\\
=\displaystyle\left|A_{\psi}\left[\partial_{b_{k}}T_{\psi}\left[f\right],b_{k}T_{\psi}\left[f\right]\right]\right|.
\end{array}
$$
We obtain  
$$
\begin{array}{lll}
\displaystyle(\int_{Spin(n)}\int_{\mathbb{R}^{+}}\left\Vert b_{k}T_{\psi}\left[f\right](a,\bullet,s)\right\Vert ^{2}\frac{da}{a^{n+1}}ds)^{\frac{1}{2}}\left\Vert \xi_{k}\widehat{f}\right\Vert\\
\\
\qquad\qquad\geq\sqrt{2^{n+1}\pi{}^{n}A_{\psi}}\left\Vert f\right\Vert ^{2}
+\sqrt{2^{n+3}\pi{}^{n}A_{\psi}}\left|\left[\partial_{b_{k}}T_{\psi}\left[f\right],b_{k}T_{\psi}\left[f\right]\right]\right|.
\end{array}
$$
Since for $f,g\in L^{2}(\mathbb{R}^{n},dV(\underline{x}))$ it holds that
\[
\left[T_{\psi}\left[f\right],T_{\psi}\left[g\right]\right]=\left\langle f,g\right\rangle 
\]
we may write 
$$
\begin{array}{lll}
\displaystyle(\int_{Spin(n)}\int_{\mathbb{R}^{+}}\left\Vert b_{k}T_{\psi}\left[f\right](a,\bullet,s)\right\Vert ^{2}\frac{da}{a^{n+1}}ds)^{\frac{1}{2}}\left\Vert\xi_{k}\widehat{f}\right\Vert\\
\\
\qquad\qquad\qquad\qquad\geq\sqrt{2^{n+1}\pi{}^{n}A_{\psi}}\left\{\left\Vert f\right\Vert ^{2}+2\left|\left\langle f_1,f_2\right\rangle \right|\right\}
\end{array}
$$
where
$$
f_1(\underline{x})=\frac{1}{A_{\psi}}\int_{Spin(n)}\int_{\mathbb{R}^{n}}\int_{\mathbb{R}^{+}}\psi^{a,\underline{b},s}(\underline{x})\partial_{b_{k}}T_{\psi}\left[f\right](a,\underline{b},s)\frac{da}{a^{n+1}}dV(\underline{b})ds
$$
and
$$
f_2(\underline{x})=\frac{1}{A_{\psi}}\int_{Spin(n)}\int_{\mathbb{R}^{n}}\int_{\mathbb{R}^{+}}\psi^{a,\underline{b},s}(\underline{x})b_{k}T_{\psi}\left[f\right](a,\underline{b},s)\frac{da}{a^{n+1}}dV(\underline{b})ds.
$$
\section{Conclusion}
In this paper, an uncertainty principle based on the continuous wavelet transform in the Clifford algebra's settings has been formulated and proved. Compared to xisting formulations in Clifford framework, the proposed uncertqinty principle here is sharper.
\section{Data availability statement}
Data sharing is not applicable to this article as no new data were created or analyzed in this study.

\end{document}